\documentclass{nature}

\usepackage{lineno}
\usepackage{amsmath}
\usepackage{amssymb}
\usepackage{color}

\usepackage{graphicx}
\makeatletter
\let\saved@includegraphics\includegraphics
\AtBeginDocument{\let\includegraphics\saved@includegraphics}
\renewenvironment*{figure}{\@float{figure}}{\end@float}
\makeatother

\newcommand{\cm}	{\mbox{cm}}

\newcommand{\kms}	{\mbox{km s}^{-1}}
\newcommand{\pc}	{\mbox{pc}}

\newcommand{\K} {\mbox{K}}
\newcommand{\kkms} {\mbox{K km s}^{-1}}
\newcommand{\vlsr} {V_\mathrm{lsr}}
\newcommand{\vsys} {V_\mathrm{sys}}
\newcommand{\mjybeam} {\mbox{mJy beam}^{-1}}

\newcommand{\au} {\mbox{au}}

\newcommand{\mas} {\mathrm{mas}}
\newcommand{\Te}         {T_\mathrm{e}}
\newcommand{\Tff}         {T_\mathrm{ff, 1.3 mm}}
\newcommand{\Thrl}         {T_{\mathrm{H30}\alpha}}
\newcommand{\tauff}         {\tau_\mathrm{ff, 1.3 mm}}
\newcommand{\tauhrl}         {\tau_{\mathrm{H30}\alpha}}

\newcommand{\dv}       {\Delta v}

\title{Dynamics of a Massive Binary at Birth}

\author{Yichen Zhang$^{1}$, Jonathan C. Tan$^{2,3}$, Kei E. I. Tanaka$^{4,5}$,
James M. De Buizer$^{6}$, Mengyao Liu$^{3}$, 
Maria T. Beltr\'an$^{7}$, Kaitlin Kratter$^{8}$,
Diego Mardones$^{9}$, Guido Garay$^{9}$}

\begin{document}

\maketitle

\begin{affiliations}
 \item Star and Planet Formation Laboratory, RIKEN Cluster for Pioneering Research, Hirosawa 2-1, Wako-shi, Saitama 351-0198, Japan
 \item Department of Space, Earth \& Environment, Chalmers University of Technology, SE-412 96 Gothenburg, Sweden
 \item Department of Astronomy, University of Virginia, Charlottesville, VA 22904-4325, USA
 \item Department of Earth and Space Science, Osaka University, Toyonaka, Osaka 560-0043, Japan
 \item Chile Observatory, National Astronomical Observatory of Japan, Mitaka, Tokyo 181-8588, Japan
 \item SOFIA-USRA, NASA Ames Research Center, MS 232-12, Moffett Field, CA 94035, USA
 \item INAF -- Osservatorio Astrofisico di Arcetri, Largo E. Fermi 5, 50125 Firenze, Italy
 \item Department of Astronomy and Steward Observatory, University of Arizona, 933 N Cherry Ave, Tucson, AZ, 85721, USA
 \item Departamento de Astronom\'ia, Universidad de Chile, Casilla 36-D, Santiago, Chile
\end{affiliations}


\begin{abstract}
Almost all massive stars have bound stellar companions, existing in
binaries or higher-order
multiples\cite{Chini12,Sana12,Peter12,Almeida17,Moe17}. While binarity
is theorized to be an essential feature of how massive stars
form\cite{Kratter10}, essentially all information about such
properties is derived from observations of already formed stars, whose
orbital properties may have evolved since birth. Little is known about
binarity during formation stages.  Here we report high angular
resolution observations of 1.3~mm continuum and H30$\alpha$
recombination line emission, which reveal a massive protobinary with
apparent separation of 180~au at the center of the massive
star-forming region IRAS07299-1651.  From the line-of-sight velocity
difference of 9.5$\:\kms$ of the two protostars, the binary is
estimated to have a minimum total mass of 18 solar masses,
consistent with several other metrics, and maximum
period of 570 years, assuming a circular orbit.
The H30$\alpha$ line from the primary
protostar shows kinematics consistent with rotation along a ring of
radius of 12~au.
The observations indicate that disk fragmentation at several hundred
au may have formed the binary, and much smaller disks are feeding the
individual protostars.
\end{abstract}

We observed infrared source IRAS07299-1651, thought to be a massive
protostar\cite{Debuizer17}, $1.68\:$kpc away\cite{Reid09}, with the
Atacama Large Millimeter/Submillimeter Array (ALMA) (Methods). At
$\sim10^4\:{\rm{au}}$ scales, the low-resolution
($0.22''\times0.15''$, i.e., $370\:\au\times260\:\au$) 1.3~mm
continuum image exhibits several stream-like structures connecting to
the central source (Figure~\ref{fig:continuum}a).
The mass of these structures $>500\:$au from the
continuum peak is $3.8-8.0\:M_\odot$ (Methods).
The high angular resolution ($35\:\mas\times29\:\mas$, i.e.,
$59\:\au\times49\:\au$) 1.3~mm continuum observation filters out
large-scale emission and resolves the central peak into two compact,
marginally-resolved sources with apparent separation of 180~au
(Figure~\ref{fig:continuum}b). The fluxes of the brighter, western
Source A and the fainter, eastern Source B are 51 and 18~mJy,
respectively.

H30$\alpha$ hydrogen recombination line (HRL) emission is detected
towards both sources, with positions and sizes coinciding closely with
the continuum emission (Figure~\ref{fig:continuum}b). 
  The strong HRL emission suggests that the small-scale 1.3~mm
  continuum has significant contribution from ionized gas free-free
  emission, in addition to dust emission (Methods).
H30$\alpha$ spectra from the continuum peak positions are shown in
Figure~\ref{fig:spectrum}.  They are well fit ($<$10\% deviation) with
Gaussians from which central velocities are determined:
$15.5\pm1.1\:\kms$ for A and $25.0\pm1.6\:\kms$ for B (Methods).
Source A's spectrum exhibits slight asymmetry, perhaps caused by small
portions of optically thick gas, or different internal velocity
components.  
Assuming the H30$\alpha$ central velocities trace the 
protostar radial velocities (Methods), the velocity difference between the 
two sources may then be due to binary orbital
motion, and can thus constrain the system mass and orbital
properties.

First assuming circular (zero eccentricity) orbits, expected to be a good
approximation for a binary forming by disk fragmentation, i.e., via
accretion of gas on near circular disk orbits\cite{Kratter10}, then
the minimum source separation is their apparent separation
$a_0=180\pm11\:\au$, with uncertainty dominated by that of source
distance\cite{Reid09}.  Combining with the projected velocity
difference 
$\Delta v=9.5\pm1.9\:\kms$, yields a maximum orbital period
$P_0=(5.7\pm1.2)\times 10^2$ years and minimum total system mass
$M_0=18.4\pm7.4\:M_\odot$. For an elliptical orbit with
  eccentricity $e$, the minimum system mass is
  $M_\mathrm{min}=M_0/(1+e)$.  The minimum mass for a bound system is
  therefore $M_0/2=9.2\pm3.7\:M_\odot$.  
For circular orbits, Figure~\ref{fig:binary} displays allowed
distributions of orbital period and system mass, showing how changes
in orbital plane inclination and position angle cause the system mass to
be $>M_0$ and the orbital period to be $<P_0$.  Supplementary
  Figure 1 shows similar distributions for elliptical orbits.  Typical
  example orbits with $e\leq0.2$ are displayed in Figure
  \ref{fig:continuum}(c).
Assuming the center of mass radial velocity of the binary
is the same as the molecular cloud systemic velocity,
we also constrain the binary mass ratio from the determined source
radial velocities (Methods). With $\vsys=16.5-18~\kms$, the probable range
for the binary mass ratio is up to $\sim0.7$
(Figure \ref{fig:binary}c).

We also use the HRL-derived free-free emission to
  independently estimate protostar masses and thus further constrain
  binary orbital properties.  We estimate free-free components in the
  1.3~mm continuum to be 39~mJy and 4~mJy in Sources A and B (Methods).  If
free-free emissions arise from regions ionized by stellar radiation,
implied zero-age main-sequence (ZAMS) masses are $12.5\:M_\odot$
(Source A) and $10\:M_\odot$ (Source B), i.e., spectral types B0.5 and
B1 (Methods), suggesting indeed two massive stars in
  formation.  However, the protostars may not yet have contracted to
the ZAMS. Then free-free emission implies masses $8-19\:M_\odot$ for A
and $7-17\:M_\odot$ for B (Methods). The concentrated HRL emission
morphology suggests the ionized gas is confined close to the
protostars, consistent with theoretical models for the above mass
estimates\cite{Tanaka16}. The total system masses from
  these estimations are also consistent with the minimum mass from
  orbital constraints.

As Figure~\ref{fig:binary} shows, for the system mass of
  $\sim22.5~M_\odot$ estimated from free-free emission based on ZAMS
  models, orbital period $P$ is $510-570\:$yr, orbital plane is close
to edge-on (inclination between orbital plane and sky plane
$i>70^\circ$), and position angle of orbital plane similar (within
$5^\circ$) to the A-B axis.  Considering uncertainties in
determination of protostellar masses ($15-36\:M_\odot$),
the orbital period can be shorter ($\sim400\:$yr) and orbital plane
inclination can be $i>50^\circ$, but orbital plane position angle
remains within $\sim15^\circ$ relative to the A-B axis.
  The ranges of orbital properties increases if elliptical orbits are
  considered (Supplementary Information).

The observed H30$\alpha$ line widths, FWHM of $39$ and $55\:\kms$
of A and B, are expected to be dominated by dynamics of
  turbulence, rotation, inflow or outflow, rather than by thermal or
pressure broadening, unlike lower frequency H$n\alpha$ $(n\gg30)$
lines\cite{Keto08}.
Velocity gradients are seen, especially around the primary
(Supplementary Figures \ref{fig:momentmap} and \ref{fig:channelmap}), indicating ordered motion of ionized
gas. To understand such motion around the primary, in each velocity
channel with H30$\alpha$ peak emission $>20\sigma$, a 2D Gaussian fit
is performed to determine the emission's centroid position (Figure
\ref{fig:centroid} and Methods).  Centroid positions show a very
organized pattern along a half ellipse with center close to the
continuum peak.  The northern half of the ellipse is blue-shifted; the
southern is red-shifted. The most blue and red-shifted
emission is at the ends of the major axis.  One way to explain such a
pattern is by an inclined rotating ring: we fit centroid positions
and intensities with such a model (Methods). The
  best fit model (Figure~\ref{fig:centroid}) has a ring with radius
  $R_\mathrm{ring}=7\pm1\:\mas$, i.e., 12~au,
rotating at velocity $V_\mathrm{rot}=21\pm2\:\kms$, corresponding to a
central mass of $6\pm2\:M_\odot$, assuming Keplerian rotation.

This dynamic mass of the primary is consistent with the
  minimum system mass constrained from orbital motion (assuming
  similar masses of the binary members).  It is somewhat smaller than
  that estimated from free-free emission
  ($12.5^{+6.5}_{-4.5}\:M_\odot$), so the rotation might be
  sub-Keplerian. Such a rotating structure is likely to either be part
  of the accretion disk that has been ionized\cite{Keto03,Keto06} or
  from a slow, rotating ionized disk wind, which have been
  seen in some other systems\cite{Guzman14,Zhang17}.
 The small-scale dust continuum emissions are found to be optically
 thick, suggesting structures with high mass surface densities of
 $\sim1\times10^2\:\mathrm{g}\:\mathrm{cm}^{-2}$ around the protostars
 (Methods), which are likely to be individual circumstellar disks.  If
 the ionized gas ring is confined within an opaque dusty disk that
 is thick and flared, due to the inclination, the front side of the
 ionized ring would be blocked by the outer part of such disk.
 This can naturally explain why only the eastern half of the ring,
 which is the far side, is seen in H30$\alpha$ emission.

The morphology and kinematics of the large sclae structures
appear complex, as illustrated by zeroth and first moment maps of
CH$_3$OH line emission (Supplementary Figure \ref{fig:infall}a).  We use a model of
rotating-infall\cite{Sakai14} to explain the kinematic features of one of the main structures
(Supplementary Information).  This model requires a central mass of
$27\pm6~M_\odot$, consistent with the minimum system mass
  derived from orbital motion and also the total protostellar mass
  estimated from free-free emission.  The radius of the centrifugal
barrier is estimated to be $840~\au$, moderately larger
  than the binary separation, as expected in disk fragmentation
  models\cite{Kratter10}.  A circumbinary disk may have formed inside
  the centrifugal barrier, which feeds one or both members of the
  binary.  However, it is difficult to separate such disk emission
  from that of the infalling streams due to projection effects. Also,
  there is no distinct kinematic signature of a circumbinary disk
  detected in CH$_3$OH emission.

Placing our results in context, so far only very few massive
protobinary systems have been identified: IRAS20126+4104 (apparent
separation of 850~au) by NIR imaging\cite{Sridharan05}, G35.20-0.74
(800 au) and NGC7538-IRS1 (430~au) by cm continuum
observations\cite{Beltran16,Beuther17}, and IRAS17216-3801 (170~au) by
NIR interferometry\cite{Kraus17}.  Only the separation information was
used to define these binary systems.  Our study is the first, to our knowledge, 
measurement of dynamical constraints on the or- bit of a forming massive binary.  
In IRAS07299-1651 we are witnessing massive binary formation and
accretion on multiple spatial scales, from infalling streams at
$1,000-10,000$~au, to formation of a massive binary system on
$100-1,000$~au scales, and to accretion disks feeding individual stars
on 10~au scales.

Overall we consider these results support a scenario of disk
fragmentation for massive binary formation\cite{Kratter10}. First,
large-scale structures are seen that are consistent with infall in the
core envelope to a central disk, and the observed separation of the
binary is moderately smaller than the inferred centrifugal
barrier. Second, the secondary-to-primary mass ratio is about 0.8
based on free-free emission
and ZAMS models, or up to about 0.7 based on the center of mass velocity,
close to asymptotic values seen in simulations of binary formation via
disk fragmentation. This is caused by the secondary growing
preferentially from a circumbinary disk, since it is further from the
center of mass. Production of near equal mass high-mass stars by
turbulent fragmentation, i.e., independent formation events that
happen to form near each other in a bound state, is unlikely, given
the rarity of massive stars. Third, only very few protostellar sources
are detected in the region (Supplementary Information), rather than a
rich cluster of forming stars, which is consistent with limited
fragmentation in Core Accretion models for massive star
formation\cite{MT03}, perhaps due to magnetic fields, radiative
heating or the tidal fields from an already formed central massive
star or binary. In this case, fragmentation to produce the binary at
its observed scale arises from gravitational instability in a massive
disk around the original primary. The current circumbinary disk could
be much less massive than this earlier disk. A much higher degree of
fragmentation is expected if the region is undergoing widespread
turbulent fragmentation and in Competitive Accretion models of massive
star formation\cite{Bonnell01}.

There are, however, caveats and open questions associated with the
disk fragmentation interpretation.  One concerns the potential
misalignment between the orbital plane and the rotational structure
around Source A. The angle between these two planes is $>54^\circ$
(Methods). Furthermore, the direction of the large scale structures
appears to be similar to that of the rotational structure around
Source A and different from that of the orbital plane.  While such
misalignment is often considered as an indicator of turbulent
fragmentation of binary formation, it may also be caused by changes in
the orientation of the angular momentum of accreting gas at these
various scales, perhaps inherited from different infalling components
of a turbulent core, where one expects substructure in the infall
envelope\cite{Myers13}. Indeed the infalling material appears highly
structured and would have different angular momentum directions, so
disk orientation should fluctuate during the formation process.
Such misalignment between circumbinary and circumstellar disks 
have been indicated by recent observations of both low and high-mass sources\cite{Kraus17,Takakuwa17}.
Future observations may test the disk fragmentation scenario by
determining whether the orbit is close to circular or not. Finally,
larger samples need to be observed with these methods to determine how
common these features are during massive star formation.

\begin{addendum}
 \item[Correspondence] Correspondence and requests for materials
should be addressed to Y.Z. (email: yichen.zhang@riken.jp).
 \item The authors thank Nami Sakai for valuable discussions.  ALMA is
   a partnership of ESO (representing its member states), NSF (USA)
   and NINS (Japan), together with NRC (Canada), MOST and ASIAA
   (Taiwan), and KASI (Republic of Korea), in cooperation with the
   Republic of Chile.  The Joint ALMA Observatory is operated by ESO,
   AUI/NRAO and NAOJ.  The National Radio Astronomy Observatory is a
   facility of the National Science Foundation operated under
   cooperative agreement by Associated Universities, Inc.
   Y.Z. acknowledges support from RIKEN Special Postdoctoral
   Researcher Program. J.C.T. acknowledges support from NSF grant
   AST1411527 and ERC Advanced Grant project MSTAR.  K.E.I.T
   acknowledges support from NAOJ ALMA Scientific Research Grant
   Number 2017-05A. 
   D.M. and G.G. acknowledge support from CONICYT project Basal AFB-170002.
 \item[Author contributions] Y.Z. led part of the ALMA observations,
   performed the data analysis, led the discussions, and drafted the
   manuscript.  J.C.T. led part of the ALMA observation, and
   participated in the discussions and drafting manuscript.
   K.E.I.T. contributed to the discussions.  The rest of the authors
   discussed the results and commented on the manuscript.
 \item[Competing interests] The authors declare that they have no
competing financial interests.
\end{addendum}

\clearpage

\begin{figure}
\begin{center}
\includegraphics[width=\textwidth]{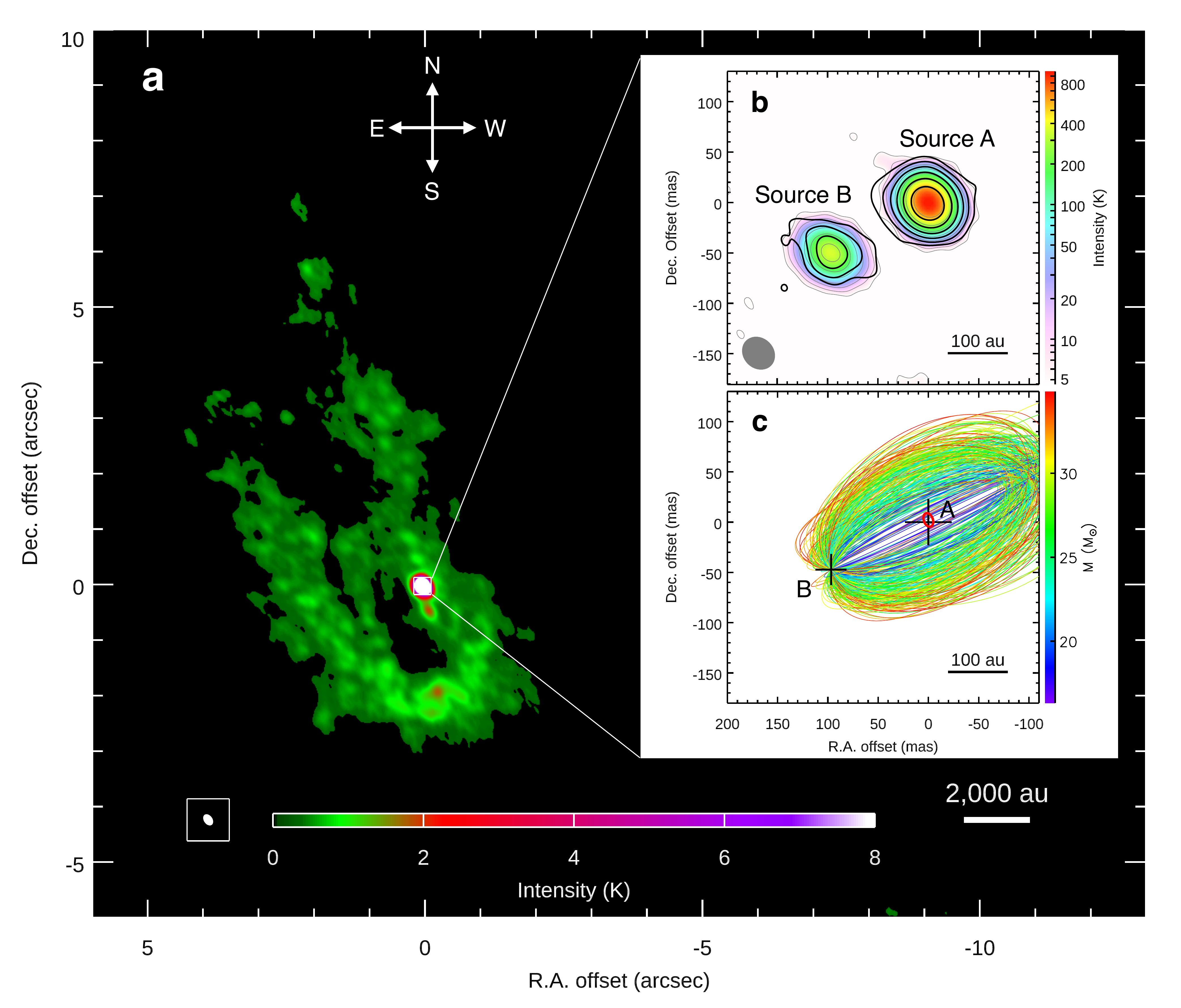}
\end{center}
\end{figure}

\clearpage

\begin{figure}
\begin{center}
\caption{
{\bf Maps of the continuum and H30$\alpha$ line emissions. (a):} The
1.3 mm continuum map is shown in color scale on the background image.
The synthesized beam (shown in box in bottom-left corner) is
$0.22''\times 0.15''$.  {\bf (b):} The 1.3 mm continuum map (color
scale and white contours) and the H30$\alpha$ line emission (dashed
orange contours) on small scales revealing a binary
  system. The continuum contour levels are 3$\sigma$, 6$\sigma$,
12$\sigma$, 24$\sigma$, ..., with $1\sigma=1.6~\K$ (0.07 $\mjybeam$).
The H30$\alpha$ line emission is integrated in the velocity range of
$-30~\kms<\vlsr<55~\kms$, and the contour levels are 5$\sigma$,
10$\sigma$, 20$\sigma$, ..., with $1\sigma=280~\kkms$ (13 $\mjybeam$
$\kms$). The synthesized beam (shown in bottom-left corner) is
$35~\mas\times 29~\mas$.  {\bf (c):} Examples of possible binary
orbits (relative orbits of Source B with respect to Source A) of
different system masses shown in color. Orbits with system mass
ranging from 5 to 35 $M_\odot$ and eccentricity ranging
from $e=0$ to 0.2 are shown.  The red ellipse around Source A is the
rotational structure fitted from H30$\alpha$ emission centroids.  The
R.A. and Dec. offsets are relative to the continuum peak position of
Source A ($7^\mathrm{h}32^\mathrm{m}09^\mathrm{s}.785,
-16^\circ58'12''.148$).}
\label{fig:continuum}
\end{center}
\end{figure}

\clearpage

\begin{figure}
\begin{center}
\includegraphics[width=0.8\textwidth]{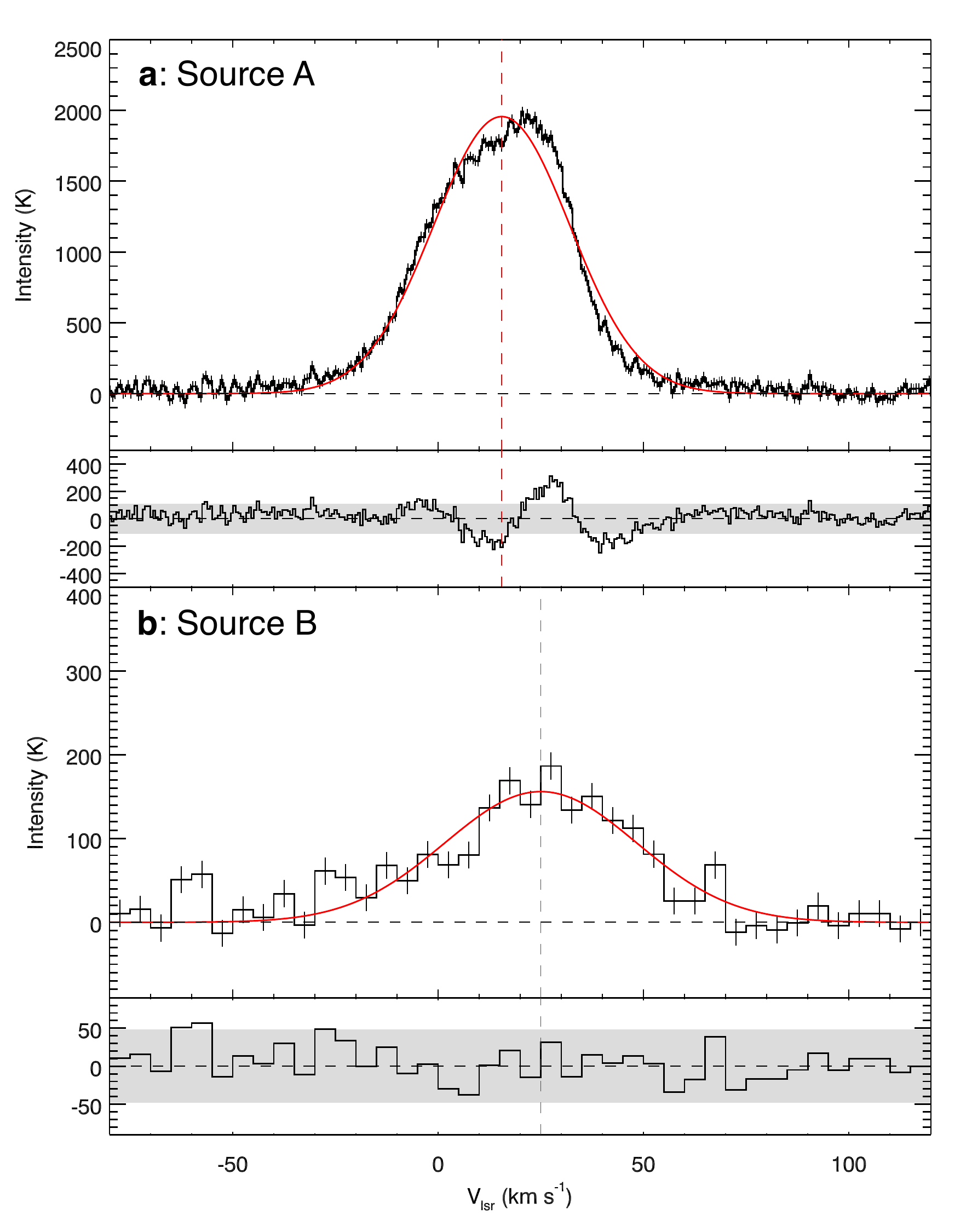}
\caption{
{\bf H30$\alpha$ line spectra at the continuum peak positions of
  Source A (panel a) and Source B (panel b)}.  The
  r.m.s. noise levels are marked by the error bar in each velocity
  channel.  The red solid curves are fitted Gaussian profiles with
the central velocities indicated by the red dashed lines.
The residual differences between the observed spectra and fitted
  Gaussian profiles are also shown below the spectra.  The shaded
  regions indicate the $3\sigma$ noise levels.}
\label{fig:spectrum}
\end{center}
\end{figure}

\clearpage

\begin{figure}
\begin{center}
\includegraphics[width=\textwidth]{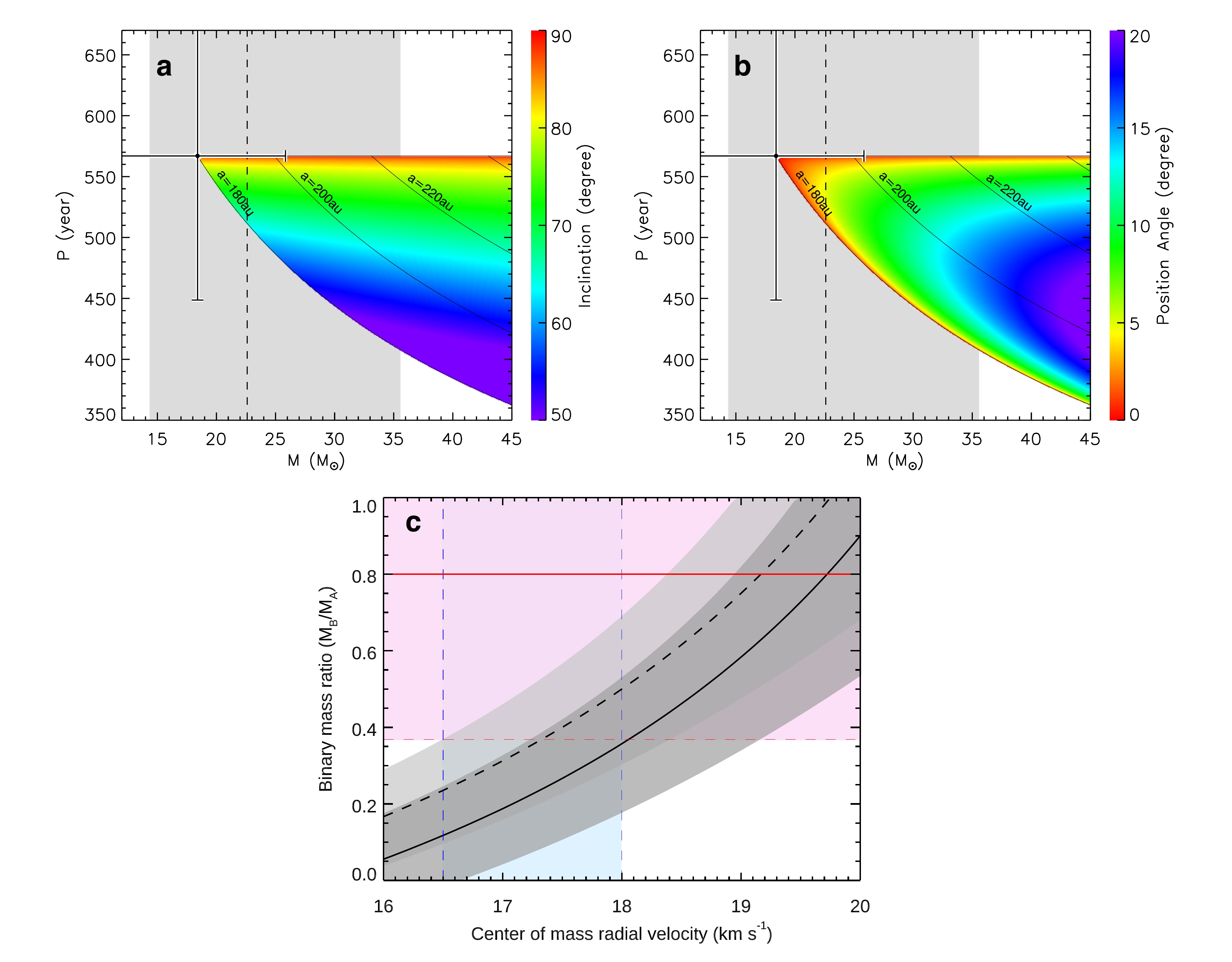}
\end{center}
\end{figure}
\clearpage

\begin{figure}
\begin{center}
\caption{
{\bf (a): The distribution of the possible binary properties in the space
  of system mass and orbital period. } The color shows the
inclination of the orbital plane relative to the plane of sky. 
Only circular
orbits are considered here.  The data point and error
  bars correspond to the system mass, orbital period and their
  uncertainties, assuming an edge-on circular orbit with the apparent
  separation of the two sources as their true separation.  The dashed
vertical line and the shaded regions indicate the system
mass and its uncertainties derived from the continuum
and H30$\alpha$ line observations. The solid lines show the locations
of orbits with different separations, as labelled.
{\bf (b):} Same as panel a, with the color showing the position angle of the intersection line
between the orbital plane and the sky plane, with respect to the
position angle of the line connecting Source A and B.
{\bf (c): The dependence of binary mass ratio ($M_B/M_A$) on the center of mass radial velocity,
calculated from the source radial velocities and uncertainties.}
The solid black curve and the dark shaded region show
this relation with $v_A=15.5\pm1.1~\kms$, and the dashed black
curve and the light shaded region show this relation with $v_A=14.5\pm 1.1~\kms$,
considering $v_A$ is likely to be more blue-shifted than the spectrum
central velocity due to the slight asymmetry in the spectrum.
The blue region between the two vertical dashed lines shows
the range of cloud systemic velocities measured from various molecular lines.
The solid red line and the red region show the mass ratio derived from
H30$\alpha$ intensities and its uncertainties.}
\label{fig:binary}
\end{center}
\end{figure}

\clearpage

\begin{figure}
\begin{center}
\includegraphics[width=\textwidth]{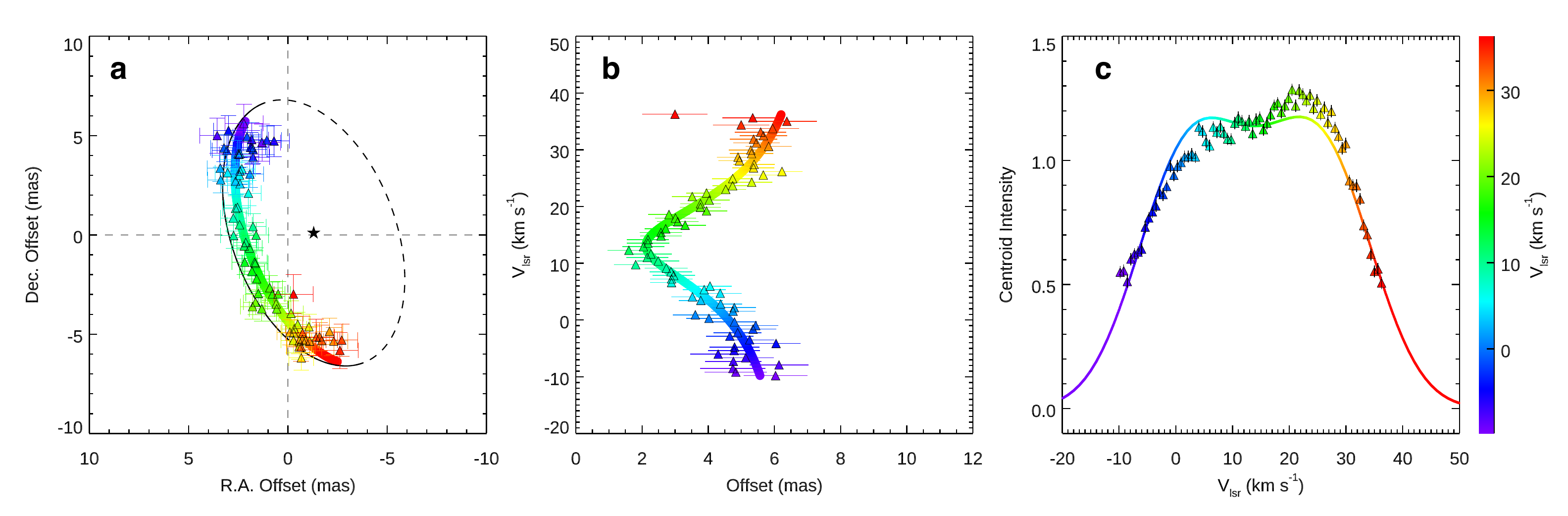}
\caption{
{\bf a) Distribution of the centroids of H30$\alpha$ emission in
  Source A in each velocity channel.}  The centroids
  (triangles with error bars) are determined by a 2D Gaussian fit to
  the H30$\alpha$ emission in each channel.  Only channels with peak
H30$\alpha$ intensities higher than 20$\sigma$
($1\sigma=1.8~\mjybeam$) are included.  The R.A. and Dec. offsets are
relative to the continuum peak position. Line-of-sight velocities are
shown by the color scale. The colored circles are the
  predicted centroid distribution of the best fit model of an inclined
  rotating ring.  The actual projected shape of the best fit ring is
  shown by the black ellipse, where the emission comes from the
  eastern half (shown with a solid line).  The center of the fitted ring
is marked by the star.  {\bf b) Distances of the centroids from the
  continuum peak position with their line-of-sight velocities,
  compared with the best-fit model.} {\bf c) Scaled
    centroid intensities compared with the prediction of the best-fit
    model.}  The meaning of the symbols in Panels (b) and (c) are
same as those in panel (a).}
\label{fig:centroid}
\end{center}
\end{figure}

\clearpage

\begin{methods}

\subsection{Observations.}
The observations were carried out with ALMA in Band 6 on April 3, 2016
with the C36-3 configuration, on Sept. 17, 2016 with the C36-6
configuration, and on Sept, 23, 2017 with the C40-9 configuration.  The
total integration time is 3, 6, and 18 min in the three
configurations.  36 antennas were used and the baselines range from
15~m to 462~m in the C36-3 configuration, 36 antennas were used and
the baselines range from 15~m to 3.2~km in the C36-6 configuration, and 40
antennas were used and the baselines range from 41~m to 12~km in the
C40-9 configuration.  J0750+1231 was used for bandpass and flux
calibration, while J0730-1141 and J0746-1555 were used as phase
calibrators.  The source was observed with single pointings, and the
primary beam size (half power beam width) was 22.9$''$.  The
data from C36-3 and C36-6 configurations were combined (referred to as
``low-resolution'' data), while the data from C40-9 configuration (referred
to as the ``high-resolution'' data) were not combined with other data,
in order to emphasize the small-scale structures.
The largest recoverable scales of the low- and
high-resolution data are about 11$''$ and 3.9$''$, respectively.

A spectral window with a bandwidth of 2~GHz was used to map the 1.3 mm
continuum.  The H30$\alpha$ line was observed with velocity resolution
of about $0.7~\kms$, and the molecular lines with about $0.2~\kms$.
The molecular lines are only detected in the low-resolution
observation, and here we only show the CH$_3$OH $4(2,2)-3(1,2)$ line
data, as other detected molecular lines such as the H$_2$CO $3(2,1)-2(2,0)$ line 
and the C$^{18}$O $(2-1)$ line show similar behaviors 
(these data will be presented in a separate paper).

The data were calibrated and imaged in CASA\cite{McMullin07}.  Self-calibration
was applied to both the continuum and line data by using the continuum
data after the normal calibration.  The self-calibration was performed
for the data of three configurations separately.  The CLEAN algorithm
was used to image the data, using robust weighting with the robust
parameter of 0.5.  The resultant synthesized beams are $0.035'' \times
0.029''$ for the high resolution continuum data, $0.035'' \times
0.030''$ for the high resolution H30$\alpha$ data, $0.22'' \times
0.15''$ for the low resolution continuum data, and $0.25'' \times
0.17''$ for the low-resolution CH$_3$OH data.

The continuum peaks of the two sources are derived to be at
$(\alpha_{2000},\delta_{2000})_\mathrm{Source~A}=(7^\mathrm{h}32^\mathrm{m}09^\mathrm{s}.786$,
$-16^\circ58'12''.146)$ and
$(\alpha_{2000},\delta_{2000})_\mathrm{Source~B}=(7^\mathrm{h}32^\mathrm{m}09^\mathrm{s}.793$,
$-16^\circ58'12''.196)$
from the high-resolution data using CASA imfit. 
There is a third continuum source at 8.5$''$ ($1.4\times 10^4~\au$) to the
northeast of the binary system, detected in both low and high
resolution data (not shown in Figure \ref{fig:continuum}).  In the
high-resolution data, the source has a peak brightness temperature of
30 K and a resolved size of about 200 au, i.e., much fainter and more
extended than the binary sources.  No H30$\alpha$ emission is detected
toward this source. 
Multiplicity in this region is discussed in Supplementary Information.

\subsection{Estimating the line-of-sight velocities of the protostars.}
We fit the H30$\alpha$ spectra at the continuum peak positions
of the two sources with Gaussian profiles to determine the central velocities.
The uncertainties are estimated by adding random noise 
(same as the r.m.s noise level of each channel)
to the fitted Gaussian profiles, repeating the fitting many times,
and then calculating the standard deviation of the fitted central
velocities.
For Source A, the spectrum deviates
slightly from a symmetric Gaussian profile, but the fractional
difference of the data from the best-fit Gaussian profile is only 7.8\%
(only the parts which have deviations $>3\sigma$ are included).  For
Source B, the deviations of the data from the Gaussian profile are all
within the $3\sigma$ level.  The determined central velocities are
$v_A=15.5\pm0.064\:\kms$ for Source A and $v_B=25\pm1.2\:\kms$ for Source B.

For Source A, we further add some perturbation on the fitted Gaussian profile to 
simulate the effects of the asymmetry on the determination of central velocity.
The perturbation has a form of sine function with a random phase,
\begin{equation}
I(v)=G(v)\left[1+A \sin\left( \frac{v}{v_0}+\phi\right)\right],
\end{equation}
where $G(v)$ is the fitted Gaussian profile.  From the residual of the
Gaussian fitting (Figure \ref{fig:spectrum}a), the perturbation
amplitude $A$ is around 0.15 and the perturbation period $v_0$ is
about $35~\kms$.  We then add this perturbation with $\phi$ randomly
distributed between 0 and $2\pi$, $A$ randomly distributed from 0.1 to
0.3, and $v_0$ randomly distributed from 25 to 50 km/s, to the
best-fit Gaussian profile, in addition to random noise, and perform
Gaussian fitting many times to determine the standard deviation of the
fitted central velocities of the simulated spectra.  The resultant
uncertainty is $1.1~\kms$, which is significantly larger than the
uncertainty $0.064~\kms$ from the pure Gaussian fitting, suggesting
the uncertainties from the asymmetry in the spectrum is the dominant
source of the error of the central velocity.  
We also perform Gaussian fitting to only the wing parts of the spectrum 
($\vlsr < 0~\kms$ or $\vlsr > 30~\kms$), and also the overall spectrum within a
radius of 0.03$''$ from the continuum peak position,
the fitted central velocities are both $14.5~\kms$. 
The differences of these measurements to the above determined central velocity are
$\sim1~\kms$.  So an uncertainty of $1.1~\kms$ for source A is
reasonable.  Therefore, we adopt $v_A=15.5\pm1.1\:\kms$ as the central
velocity of Source A spectrum.  Due to relatively low S/N radio,
Source B's spectrum appears to be quite symmetric.  However, if we
allow similar uncertainty from potential asymmetry in Source B's
spectrum, the central velocity of Source B should be
$v_B=25\pm1.6\:\kms$.  Here the uncertainty $1.6~\kms$ combines the
uncertainty from possible asymmetry ($1.1~\kms$) and the uncertainty
from original Gaussian fitting ($1.2~\kms$) following error
propagation.  The velocity difference between the two central
velocities is then $\Delta v\equiv |v_A-v_B| =9.5\pm1.9~\kms$.

We assume that the measured H30$\alpha$ central velocities
  are a good measure of the true source velocities.  Previous
  arcsec-resolution mm HRL observations towards massive protostars
  provide different results about whether the HRL central velocities
  can trace the source velocities.  Some studies show that the central
  velocities of HRLs may be offset from the molecular gas velocities\cite{Klaassen18}
  by $2-20~\kms$ or different HRLs of same
  source have central velocities different\cite{ZhangC14} by $\sim5~\kms$, 
  while some studies show that multiple
  HRLs have consistent central velocities, which are consistent with
  the source velocity based on modeling\cite{Guzman14}.  However,
  there was no mm HRL observation with a spatial resolution $<100~\au$
  previously.  With such high resolution, the HII regions in this
  source are still not resolved, suggesting an extremely early nature
  of the HII regions. As we show below, the H30$\alpha$ emission
  mostly traces the disk with motions dominated by the disk rotation,
  rather than outflow which can exhibit more complicated velocity
  structures.  In such a case we expect the central velocities of the
  H30$\alpha$ lines can better trace the source velocities.  Previous
  single dish observations of H110$\alpha$ line towards these sources\cite{Araya07}
  show a central velocity of $14.6\pm1.6~\kms$, which is consistent
  with the H30$\alpha$ central velocity of Source A, which should
  dominate the HRL emissions in low-resolution observations.  In
  addition, in this source, the cloud velocities ($16.5-18~\kms$; see below), lie
  between the determined velocities of Sources A and B,
  which is not expected if there are large offsets between the source
  velocities and the central velocities of H30$\alpha$ spectra.

We note that, since the source A spectrum shows stronger
red-shifted emission than blue-shifted emission, the true source velocity is more
likely to be blue-shifted compared to the fitted central velocity.
This indicates the velocity difference between Source A and B is more likely to be even higher
than estimated. Since we use the velocity difference to derive a
minimum mass of the binary system (see main text), the mass
constraints are rather robust even considering such possible offset in
velocity determination.

\subsection{Constraining the binary mass ratio from the source radial velocities.}
Previous single dish observations of molecular lines\cite{Araya07,Liu10} showed
systemic velocities of the surrounding molecular gas are about 16.5 to 17 $\kms$.
Gaussian fitting to the spectra of the CH$_3$OH $4(2,2)-3(1,2)$, 
H$_2$CO $3(2,1)-2(2,0)$, and C$^{18}$O $(2-1)$ lines in our ALMA data
give systemic velocities of $17-18~\kms$.
These spectra are averaged with a radius of 3$''$
from the central sources with our most compact configuration ALMA data.
If we assume the radial velocity of the center of mass
of the binary system is the same as the cloud systemic velocity ($16.5-18~\kms$), we
can estimate the binary mass ratio from the radial
velocities of the members $v_A=15.5\pm 1.1~\kms$ and $v_B=25\pm 1.6~\kms$.  As Figure
\ref{fig:binary}(c) shows, the secondary-to-primary mass ratio
ranges from $0.12\pm0.13$ with a center of mass radial velocity of
$v_\mathrm{CM}=16.5~\kms$, to $0.36\pm 0.18$ with
$v_\mathrm{CM}=18~\kms$. As discussed above, the radial velocity of
Source A is more likely to be blue-shifted compared to the fitted central velocity
$v_A=15.5~\kms$,
which suggests that the mass ratio is more likely to be higher than estimated above.
For example, with $v_A=14.5\pm 1.1~\kms$ (e.g., fitting from the
overall spectrum of the source), the mass ratio ranges
from $0.24\pm0.14$ with a center of mass radial velocity of
$v_\mathrm{CM}=16.5~\kms$, to $0.50\pm 0.19$ with
$v_\mathrm{CM}=18~\kms$.

\subsection{Estimating contributions of free-free emission and dust emission to the observed 1.3 mm continuum emission.}
Toward a massive star-forming region, the observed 1.3 mm continuum
emission may contain both free-free emission from ionized gas around
the massive protostar and dust emission, e.g., from dense molecular
gas components.  The same ionized gas also emits Hydrogen recombination lines (HRLs). 
Therefore, we first use the observed continuum-subtracted intensity of the
H30$\alpha$ HRL to estimate the level of free-free continuum emission, and then
estimate the level of the dust continuum emission by subtracting the
free-free component from the observed 1.3 mm continuum emission.

Assuming both the H30$\alpha$ line and the 1.3 mm free-free emission
are optically thin, and under the condition of local thermal
equilibrium (LTE; which is indicated by the approximately Gaussian profiles of the
observed H30$\alpha$ spectra), the ratio between the H30$\alpha$ peak
intensity and the 1.3 mm free-free emission is (according to
eqs. (10.35), (14.27) and (14.29) of ref. \citenum{Wilson13})
\begin{equation}
\frac{\Thrl}{\Tff}=4.395\times 10^6 \left(\frac{\Te}{\K}\right)^{-1} \left(\frac{\dv}{\kms}\right)^{-1}  \left[1.5\ln\left(\frac{\Te}{\K}\right)-8.443\right]^{-1}
\left[1+\frac{N\left(\mathrm{He}^+\right)}{N\left(\mathrm{H}^+\right)}\right]^{-1},
\end{equation}
where $\Te$ is the electron temperature in the ionized gas and $\dv$ is
the line width (FWHM) of the H30$\alpha$ line.  The
last term is caused by the fact that He$^+$ contributes to the
free-free emission but not to the HRL, and typically
$N\left(\mathrm{He}^+\right)/N\left(\mathrm{H}^+\right)=0.08$ 
(ref. \citenum{Shaver83}).
Assuming a typical ionized gas temperature of $\Te=8000~\K$
(e.g., ref. \citenum{Keto08}), 
for Source A, with $\dv=39~\kms$ (from Gaussian fitting of the spectrum),
we obtain $\Thrl/\Tff=2.6$, which converts to a free-free emission
contribution of about 77\% of the total 1.3~mm continuum, based on the
observed line-to-continuum emission ratio of 2.  For Source B, with
$\dv=55~\kms$ (from Gaussian fitting of the spectrum), we obtain
$\Thrl/\Tff=1.8$, which converts to a free-free emission contribution
of about 22\% of the total 1.3 mm continuum, based on the observed
line-to-continuum emission ratio of about 0.4.  These results suggest that,
for Source A, with a total continuum flux of 51 mJy, the dust
continuum emission and the free-free emission are 12 mJy and 39 mJy,
respectively, and for Source B, with a total continuum flux of 18 mJy,
the dust continuum and free-free emissions are 14 mJy and 4 mJy,
respectively.
Panels (a) and (b) of Supplementary Figure \ref{fig:lowmass} show the dependence
of the estimated free-free emission on the assumed ionized gas temperature $\Te$.
For Source A, the contribution of the free-free emission to the total 1.3 mm continuum
ranges from about 50\% at $\Te=6000~\K$ to 100\% at $\Te=10000~\K$,
which is thus the upper limit for the ionized gas temperature in this source.
For Source B, the free-free contribution ranges from 16\% at $\Te=6000~\K$
to 40\% at $\Te=12000~\K$. 
However, as we will show below, such changes of the free-free emission do not
affect the mass estimations of the protostars very much.

For optically thin emissions, $\Thrl=\Te\tauhrl$ and $\Tff=\Te\tauff$.
For Source A, the possible maximum $\Tff$ is the measured continuum
intensity $1000~\K$.  With the measured $\Thrl=2000~\K$, and assumed
$\Te=8000~\K$, the optical depths of the free-free emission and
H30$\alpha$ emission are estimated to be $\tauff<0.13$ and
$\tauhrl=0.25$.  Similar considerations gives $\tauff=0.05$ and
$\tauhrl=0.02$ for Source B.  Therefore the optically thin assumptions
are indeed reasonably valid.  Without additional observations of the
continuum emission at other frequencies, it is difficult
to separate the components of the free-free emission and dust
continuum emission more accurately.

\subsection{Estimating protostellar masses from free-free emission.}
The derived free-free emission fluxes (39 mJy for Source A and 4 mJy
for Source B) can be used to constrain the masses of the protostars if
these are from locally photo-ionized regions.  We follow the method of
ref. \citenum{Schmiedeke16}, which assumes a spherical ionized region
with a uniform electron density. Hydrogen-ionizing photon rates of
$1.5\times 10^{46}~\mathrm{s}^{-1}$ and $1.7\times
10^{45}~\mathrm{s}^{-1}$ are obtained for Source A and B,
respectively, assuming the same electron temperature of $\Te=8000~\K$.
For an ionized region of 0.03$''$ (50 au), which is the size of the
resolution beam, the electron densities are estimated to be $1.6\times
10^7~\cm^{-3}$ and $5.5\times 10^6~\cm^{-3}$ for Source A and B,
respectively.  Then their emission measures are $EM=6.7\times 10^{10}$
and $7.4\times 10^9~\pc~\cm^{-6}$.  These values will be higher if the
size of the ionized regions are even smaller.  For ZAMS stars, these
estimated ionizing photon rates correspond to stellar masses of about
12.5 $M_\odot$ and 10 $M_\odot$, i.e., spectral types B0.5 and B1
(ref. \citenum{Davies11,Meynet00,Lanz07,Mottram11}).  The luminosities
of $12.5\:M_\odot$ and $10\:M_\odot$ ZAMS stars are
$1\times10^4\:L_\odot$ and $5\times10^3\:L_\odot$ (ref
\citenum{Davies11}), whose sum is consistent with the
$(1-4)\times10^4\:L_\odot$ estimate for the total bolometric
luminosity of this system\cite{Debuizer17}.  For protostars yet to
reach the main-sequence, e.g., due to different accretion histories,
the same photo-ionizing rates can correspond to a wider range of
protostellar masses. According to protostellar evolution calculations
with various accretion histories from different initial and
environment conditions for massive star
formation\cite{Tanaka16,Zhang18}, such ionizing photon rates
correspond to protostellar mass ranges of $8-19~M_\odot$ for Source A,
and $7-17~M_\odot$ for Source B. We use these ranges as the
uncertainties of the mass estimation from the free-free emission.  As
Panels (c) and (d) of Supplementary Figure \ref{fig:lowmass} show, the estimated
ionizing photon rates and the stellar masses only have very weak
dependences on the assumed ionized gas temperature $\Te$, which lead
to uncertainties much smaller than that brought about by different
protostellar evolution histories. If dust grains survive in the
ionized region, then the ionizing photon rates and stellar masses
derived above are likely to be lower limits, due to absorption of
Lyman continuum photons by the dust.  However, the total bolometric
luminosity of this system\cite{Debuizer17} ($(1-4)\times
10^4~L_\odot$) limits the masses of the two
protostars $< 20~M_\odot$ according to both ZAMS models\cite{Davies11}
and protostellar evolution models\cite{Zhang18}.  Note
  that the ratio between the derived ZAMS masses of the two sources is
  0.8, higher than that constrained from the source radial
  velocities and the cloud systemic velocity ($<0.50\pm0.19$).
  However, considering the large uncertainties in mass determination
  from the H30$\alpha$ intensities, the possible range of the mass
  ratio constrained from the H30$\alpha$ intensities has significant overlap with
  the mass ratio constrained from the center of mass velocity analysis (Figure
  \ref{fig:binary}c).  Possible differences between the mass ratio
  determined by these two methods may be caused by the uncertainties
  in estimating ionizing photon rates from the H30$\alpha$ emissions, 
  the uncertainties of different ZAMS or
  protostellar models, and/or possible velocity difference between the
  center of mass velocity of the binary and the cloud velocity.  For a
  mass ratio of 0.8, the center of mass radial velocity is
  $19.7\pm0.9~\kms$, which is offset from the observed cloud velocity
  by $\sim 1~\kms$, which may be possible for star formation from a
  turbulent clump with about this level of velocity dispersion.
  
\subsection{Estimating ambient gas mass from dust continuum emission.}
The derived dust continuum fluxes can be used to constrain the mass of
the surrounding gas of the protostars.  For the large scale ($10^4$~au)
dust continuum emission, we can assume the dust is optically thin at
1.3~mm and well-mixed with the gas.  The gas mass can then be estimated with
the equation
\begin{equation}
M=\frac{D^2 F_\mathrm{dust,1.3mm}}{\kappa_\mathrm{1.3mm} B(T_\mathrm{dust})},
\end{equation}
where $D=1680$ pc is the distance to the source,
$F_\mathrm{dust,1.3mm}$ is the estimated dust continuum emission flux,
and $B(T)$ is the Planck function.  For $T_\mathrm{dust}=50-100~\K$,
which is consistent with the results of dust continuum radiative
transfer simulations for massive star formation\cite{Zhang14} and
observations\cite{Beltran18}, and an opacity of
$\kappa_\mathrm{1.3mm}=0.00899~\mathrm{cm}^2~\mathrm{g}^{-1}$ with a
standard gas-to-dust mass ratio of 100 included
(ref. \citenum{Ossenkopf94}), which is suitable for dense cores in
molecular clouds, the mass of this ambient gas is estimated to be
$3.8-8~M_\odot$ from the measured total flux of $390$ mJy (excluding
the emission within the central 0.3$''$ radius). This is likely to be
a lower-limit due to the spatial filtering of the extended emission in
the interferometric observation.

For the small scale (100~au) continuum emission around the two
protostars, the dust continuum emission is likely to be optically
thick.  According to radiative transfer simulations for massive star
formation, the dust temperature within about 100~au from massive
protostars of $10-12~M_\odot$ can be a few hundreds of K
(ref. \citenum{Zhang14}).  The measured continuum peak brightness
temperatures are 1000~K and 360~K for Source A and B, respectively,
and considering the above derived fractions of 23\% and 78\% for dust
emission in the two sources, the dust emission brightness temperature
of Source A and B should be about 230~K and 280~K, respectively.
Comparing these estimated dust brightness temperatures with the
expected dust temperatures from the theoretical models, suggests
that the 1.3~mm dust emission at 100~au scale is likely to be
optically thick.  In fact, the radiative transfer simulations show
that typical optical depths of accretion disks around massive
protostars from 10 to 100~au are about $0.5-2$
(ref. \citenum{Zhang13}).  If assuming a dust optical depth of
$\tau=1$ at 1.3 mm, with the same opacity
$\kappa_\mathrm{1.3mm}=0.00899~\mathrm{cm}^2~\mathrm{g}^{-1}$, the
column density of the gas surrounding the protostar is about
$1.1\times 10^2~\mathrm{g}~\mathrm{cm}^{-2}$, which corresponds to a
mass of 0.1~$M_\odot$ within a radius of 50~au.

\subsection{Determining the H30$\alpha$ emission centroids.}
For the velocity channels with peak H30$\alpha$ intensities
$>20\sigma$ ($1\sigma=1.8\:\mjybeam$ for a velocity channel width of
0.63~$\kms$), we fit the continuum-subtracted
H30$\alpha$ images with Gaussian ellipses to determine the emission
centroid position of Source A at each velocity.  During the Gaussian
ellipse fitting, a region with a radius of 50~mas centered at Source B
is masked to exclude influence from this source.  The accuracy of the
centroid position is affected by the signal-to-noise ratio (S/N) of
the data, described by the following relation\cite{Zhang17,Condon97} 
  $\Delta \theta_\mathrm{fit}=\theta_\mathrm{beam}/(2~\mathrm{S/N})$,
  where $\theta_\mathrm{beam}$ is the resolution beam size, for which
  we adopt the major axis of the resolution beam
  $\theta_\mathrm{beam}=35~\mathrm{mas}$ (59~au).  The phase noise in
  the passband data also introduces an additional error to the
  centroid positions through passband calibrations\cite{Zhang17}.  The
  phase noise in the passband calibrator J0750+1231 is found to be
  $\Delta \phi=5.3^\circ$ after smoothing of 4 channels. Such
  smoothing is the same as that used in deriving the passband
  calibration solutions.  The additional position error is $\Delta
  \theta_\mathrm{bandpass}=
  \theta_\mathrm{beam}(\Delta\phi/360^\circ)$.  And the uncertainties
  in the centroid positions are $\Delta
  \theta_\mathrm{centroid}=\sqrt{\Delta \theta_\mathrm{fit}^2+\Delta
    \theta_\mathrm{bandpass}^2}$.  The positions of these centroids
and their uncertainties are shown in Figure~\ref{fig:centroid}.

\subsection{Model fitting of the H30$\alpha$ emission centroids.}
We fit the H30$\alpha$ emission centroids of Source A with a model of
a rotating ring, described by seven free parameters: the radius of the
rotating ring $R_\mathrm{ring}$; its systemic velocity
$V_\mathrm{sys}$; its rotation velocity $V_\mathrm{rot}$; the
position of the ring center ($x_0,y_0$) with respect to the continuum
peak; the inclination of the ring with respect to the line of sight
$\theta$; and the position angle of the projected major axis of the
ring relative to north $\psi$.  In the model, we assume only half
of the ring is emitting H30$\alpha$ emission.  For each set of the
parameters, we convolve the half-ring structure with a Gaussian
profile of FWHM of 19~$\kms$ (for thermal broadening of ionized gas
with $\Te=8000~\K$; see main text) in velocity space and
convolve with the resolution beam ($35''\times 30''$) in position
space to build a simulated data cube.  We then perform 2D Gaussian
fitting to the channel maps of the simulated data cube to obtain the
model centroid positions and intensities.  The best-fit model was
obtained by varying the input parameters within reasonable ranges, and
minimizing the value of
\begin{eqnarray}
\chi^2 & = & \frac{w_\mathrm{pos}}{N}\sum \frac{(x_\mathrm{model}-x_\mathrm{centroid})^2+(y_\mathrm{model}-y_\mathrm{centroid})^2}
{\Delta \theta_\mathrm{centroid}^2}\\
 & + & \frac{w_\mathrm{int}}{N}\sum \frac{(I_\mathrm{model}-I_\mathrm{centroid})^2}{\sigma_\mathrm{centroid}^2},
\end{eqnarray}
where $(x_\mathrm{model},y_\mathrm{model})$ and
$(x_\mathrm{centroid},y_\mathrm{centroid})$ are the positions of the
model centroids and observed centroids at each velocity channel,
$\Delta \theta_\mathrm{centroid}$ is the uncertainty of the determined
centroid position, $I_\mathrm{model}$ and $I_\mathrm{centroid}$ are
the normalized centroid intensities in the model and observation,
$\sigma_\mathrm{centroid}$ is the normalized observed intensity noise.
The summation is over all the possible velocity channels, with $N$
being the total number of the channels.  Since we only focus on the
geometry and kinematics of the rotating structure and do not attempt
to reproduce the line profile, we only include the second part in the
$\chi^2$ to constrain the line width rather than the detailed line
profile.  Therefore, we use weights of $w_\mathrm{pos}=0.9$ and
$w_\mathrm{int}=0.1$ in the fitting.

The fitting gives $R_\mathrm{ring}=7\pm 1~\mas$, $V_\mathrm{sys}=14\pm
1~\kms$, $V_\mathrm{rot}=21\pm 2~\kms$, 
$x_0=-1.3^{+0.6}_{-0.1}~\mas$,
$y_0=0.1^{+0.4}_{-0.6}~\mas$,
$\theta=40^{+4}_{-8}~\mathrm{deg}$, and
$\psi=19\pm 6~\mathrm{deg}$.  The best model has
$\chi^2_\mathrm{min}=1.9$.  The uncertainty of each parameter is
estimated using the parameter range of models with $\chi^2\leq 3$
while keeping other parameters unchanged.  The rotation velocity and
radius correspond to a dynamical mass of $6\pm 2~M_\odot$, if assuming
Keplerian rotation.
From the current data, it is difficult to further constrain the radial motion
of the ring structure in addition to its rotation.

We emphasize that this model is designed to be exemplary
and illustrative, with an idealized setup with minimum number of parameters. The goal
of this model is to show that the H30$\alpha$ emission can be explained by
rotation of a disk at a radius of about 12 au in Source A. Other parameters estimated
from this model fitting are less robust. For example, if the other side of the ring is not 
completely blocked as we assumed but only extincted to some level, the ring should 
have a higher inclination angle $\theta$ than we currently estimated. In such a case, the 
central mass would also be higher.
In our simple model, we also assumed the H30$\alpha$ emission comes from a thin
annulus, however, it is also possible that it could emerge from a broader
range of disk radii extending further inward.  However, at positions
closer to the protostar, we do not detect higher-velocity emissions,
which makes it impossible to explore the rotation velocity profile with
radius to confirm whether it is Keplerian rotation or
not\cite{Sanchez13,Ilee16}.  It is possible that the inner region of
the disk that would have even higher rotation velocities is also blocked by
the outer part of a flared opaque dusty disk.

\subsection{Estimating the angles between the orbital plane,
the plane of the rotational structure around Source A, and the large scale streams.}
The position angle of the rotational structure around
Source A ($\psi_\mathrm{ring}\approx 20^\circ$ with respect to north)
is almost perpendicular to the position angle $\psi_\mathrm{orbit}$ of
the binary orbital plane (close to the direction of the line
connecting the two sources). Considering the inclinations of the
rotational structure around Source A and the orbital plane, 
the angle between the two planes (i.e., the angle between the angular momentum
directions of the orbital motion and the rotation around Source A) is 
\begin{eqnarray}
\cos\alpha & = &\cos\theta_\mathrm{ring}\sin i_\mathrm{orbit}\sin\psi_\mathrm{ring}\sin\psi_\mathrm{orbit}\nonumber\\ 
& + & \cos\theta_\mathrm{ring}\sin i_\mathrm{orbit}\cos\psi_\mathrm{ring}\cos\psi_\mathrm{orbit}\nonumber\\
& + & \sin\theta_\mathrm{ring}\cos i_\mathrm{orbit},
\end{eqnarray}
where $\theta_\mathrm{ring}=40^{+4}_{-8}~\mathrm{deg}$ is the
inclination of the ring structure to the line of sight,
$i_\mathrm{orbit}>50^\circ$ is the inclination of the orbital plane
with the plane of sky, $\psi_\mathrm{ring}=19\pm 6~\mathrm{deg}$ is
the position angle of the ring structure around Source A, and
$\psi_\mathrm{orbit}$ is the position angle of the orbital plane,
which is within $15^\circ$ from the direction of the line connecting
the two sources, i.e., $101^\circ<\psi_\mathrm{orbit}<131^\circ$.
From these values, we estimate that $\alpha>54^\circ$.
If we assume that the binary orbital plane has the same inclination and 
  position angle as the Source A disk (inclination of $\sim 40^\circ$ and position angle of $\sim 20^\circ$),
  the dynamical mass of the system would be $> 45~M_\odot$ with $e < 0.9$, and $> 156~M_\odot$ with $e < 0.5$. 
  Therefore it indeed requires an unreasonable system mass, which is much larger than those estimated from the 
  HRL intensities, disk model and infall model) for the orbital plane to be aligned with the disk plane of Source A.
  However, we note that these results can be affected by the simplified and
  idealized model fitting for the rotational structure around Source
  A. Thus we do not consider this estimation to be very robust.
  The observed direction of the large scale structures
  appears to be similar to that of the rotational structure around
  Source A. In the mid-IR, this source shows $10^4~\au$-scale emission
  elongated in the NW-SE direction\cite{Debuizer17}, indicating an
  outflow cavity (shared by the binary) has formed in the direction
  perpendicular to the large-scale infalling streams.  However, no
  molecular outflows are reported so far from this source to confirm the
  outflow direction.

\subsection{Data avalability.}
This paper makes use of the following ALMA data: ADS/JAO.ALMA\#2015.1.01454.S, 
ADS/JAO.ALMA\#2016.1.00125.S. 
The data that support the plots within this paper and other findings of this study 
are available from the corresponding author upon reasonable request.

\end{methods}

\clearpage

\subsection{Supplementary Information}

\subsection{Binary properties if allowing for non-circular orbits.}
If elliptical orbits are considered, then the ranges of allowed binary
orbital properties increase compared to those assuming circular orbits
(Figure 3 of main paper). The minimum mass of the system for all
possible elliptical orbits with an eccentricity of $e$ is
$M_\mathrm{min}=M_0/(1+e)$, where $M_0=18.4\pm7.4~M_\odot$ is the
reference system mass assuming an edge-on circular orbit with the
apparent separation of the two sources as their true separation.  This
provides a lower mass limit of $M_0/2=9.2\pm3.7~M_\odot$ for the
system to be bound.  The total mass of the binary estimated from the
free-free emissions is $22.5^{+13.5}_{-7.5}~M_\odot$, which is
consistent with this minimum mass.

Supplementary Figure \ref{fig:binary1} shows the distribution of
possible binary properties in the space of total system mass and
orbital period, with eccentricities $0\leq e<0.9$.  For orbits with
low eccentricities $e<0.2$, with the estimated system mass of
$22.5^{+13.5}_{-7.5}~M_\odot$, the semi-major axis of the orbit is
around $200\:\au$, the orbital period is about $400\:\mathrm{yr}-
1000\:\mathrm{yr}$, and the inclination angle is likely to be
$i>50^\circ$. The position angle of the orbital plane is still in a
similar direction as the two sources.  Some typical orbits (relative
orbits of Source B with respect to Source A) with $e<0.2$ are shown in
Figure 1(c) of main paper. The constraints are weaker if the orbit has
a higher eccentricity. With the estimated system mass of
$22.5^{+13.5}_{-7.5}~M_\odot$, the minimum orbital period is about 300
years with a semi-major axis of 130 au, and the maximum orbital period
is about 10,000 years for $e<0.8$ with a maximum semi-major axis of
about 1,000 au.

\subsection{The large-scale gas structure and possible circumbinary disk.}
The morphology and kinematics of the large-scale stream-like
structures appear complex, as shown by the zeroth and first moment
maps of CH$_3$OH line emission (Supplementary Figure
\ref{fig:infall}a).  The CH$_3$OH emission peak is offset from the
continuum peak of the central region, extending about $0.5''$ (840~au)
in the south. A velocity gradient is seen in this elongated structure
with the most red-shifted velocity associated with the southern tip. In
order to understand such kinematics, the position-velocity diagram is
made along a cut with a position angle of $20^\circ$, passing through
the continuum peak and the elongated structure (Supplementary Figure
\ref{fig:infall}b).  The fact that the highest velocity and the
highest velocity gradient both appear at a position $0.5''$ offset
from the center can be naturally explained by infalling motion with
angular momentum conserved.  In such motion, the material reaches its
maximum rotational velocity at the radius of centrifugal barrier where
all the kinetic energy turns into rotation\cite{Sakai14}.

We construct a model to explain the kinematic features seen in
Supplementary Figure \ref{fig:infall}(b).  In this model, the material
is infalling and rotating with angular momentum conserved.  For
simplicity, the motion of material is assumed to be in a plane viewed
with zero inclination with the line of sight.  This is supported by
the fact that the large-scale structure seen in the 1.3 mm continuum
and molecular line emission is missing in the mid-IR\cite{Debuizer17},
indicating high extinction along this structure, which is natural for
a plane of accretion that is close to edge-on.  The motion can be
described by
\begin{eqnarray}
v_\varphi & = & -v_0 \frac{r_0}{r},\\
v_r & = & -v_0\frac{\sqrt{r_0(r-r_0)}}{r}.
\end{eqnarray}
Such motion conserves both angular momentum and kinetic energy. Here
$r_0$ is the innermost radius that such infalling gas can reach with
angular momentum conserved, where all the kinetic energy turns into
rotation (i.e., the centrifugal barrier), and $v_0$ is the rotational
velocity at $r_0$.  The shape of the structure is assumed to follow
the trajectory of such motion, which is a parabola, i.e.,
\begin{equation}
r = \frac{2r_0}{1+\cos\varphi}.
\end{equation}
Here the observer is at the direction of $\varphi=\pi/2$.  The offset
and line-of-sight velocity are
\begin{eqnarray}
x & = & -r\cos\varphi,\\
\vlsr & = & V_\mathrm{sys}-v_r\sin\varphi-v_\varphi\cos\varphi,
\end{eqnarray}
where the signs are selected to be consistent with the figure.
In the model, $r_0=0.5''$ (840 au), 
$V_\mathrm{sys}=15~\kms$, $v_0=7.5~\kms$.
The central mass is connected to the radius and velocity of
the centrifugal barrier by
\begin{equation}
v_0=\sqrt{\frac{2GM}{r_0}},
\end{equation}
which leads to a central mass of $27~M_\odot$.  If we use the width of
the emission at $v_\mathrm{lsr}=V_\mathrm{sys}+v_0=22.5~\kms$ in the
PV diagram as the uncertainty of $r_0$ (about $~20\%$), and assuming
velocity uncertainty of about $0.5~\kms$, the uncertainty of the mass
estimation is then about 24\%.  This mass is close to the total mass
of $22.5^{+13.5}_{-7.5}~M_\odot$ of the binary estimated from the
free-free fluxes.  It is also consistent with the
minimum mass of $18.4\pm 7.4~M_\odot$ derived from the orbital motion
for low eccentricity orbits, and the minimum mass of $9.2\pm
3.7~M_\odot$ for all possible bound orbits.

This example model is meant to be illustrative of the possibility that
the large-scale stream is infalling with rotation. Note, we do not try
to explain all the observed structures, which is difficult considering
the clumpy distribution of the material, possible projection effects,
and the filtering of large-scale emission in our interferometric
observations.  Note that in the above model, the systemic velocity
(the radial velocity of material at infinite distance) is 15~$\kms$,
slightly offset from the estimates of the cloud systemic velocity
($16.5-18~\kms$; see Methods).  Such an offset may arise if the
envelope has substructures with slightly different initial velocities.
The highest velocity that the infalling stream reaches is $\sim
23~\kms$, offset from the overall systemic velocity of $16.5-18~\kms$
of the cloud by $5-6.5~\kms$. If we adopt this velocity offset for
$v_0$, the central mass is $12-17~M_\odot$, which is still consistent
with the minimum mass derived from the orbital motion. This mass is a
lower limit if we further consider an inclination that is not fully
edge-on.  Besides the elongated emission peak, other more extended
structures with slight velocity gradients are seen in the PV diagram,
which may be part of other infalling streams with similar motions. The
velocity gradients are naturally low if they are still distant from
the central source.

In the model presented above, we consider that all the emission (even
those toward the central sources in projection) is associated with
infalling envelope structures. However, we cannot rule out the
possibility that some of the emission toward the central sources in
projection comes from an embedded circumbinary disk.  We perform a
simple estimation of the mass of any potential circumbinary material
that may be present. We first convolve the high-resolution continuum
emission of only the two sources with the resolution beam of the
low-resolution continuum data, and then subtract the convolved map
from the low-resolution continuum data. Since the continuum emission
of the two sources in the high-resolution data is dominated by the
free-free and dust emission in the immediate vicinities of the two
sources, the residual continuum emission should only contain dust
emission from surrounding materials. Within a radius of $0.5''$
($840~\au$) from Source A, the total flux of the residual continuum
emission is 12 mJy.
Assuming a dust temperature of $50 - 500$ K (as expected from
radiative transfer models of massive protostars), the continuum flux
of 12 mJy corresponds to a mass of $0.26 - 0.023~M_\odot$.  However,
it is hard to confirm that this mass is indeed in a circumbinary
disk. It may be belong to the parts of the infalling streams that
happen to be close to the source in projection. As the
position-velocity diagram shows, there is no distinct kinematic
feature toward the position of the protostars to separate the
circumbinary disk from the larger streams.  We also note that there is
no guarantee that disk fragmentation produces a substantial
circumbinary ring. A substantial circumbinary disk may or may not
exist after the fragmentation according to
simulations\cite{Krumholz07,Kratter10}.

\subsection{The low-mass protostellar sources in the region.}
We performed a search for low-mass protostellar sources in the region
using the high resolution continuum images.  We identify the sources
with maximum intensities $>5\sigma$ and sizes of regions with
intensities $>4\sigma$ larger than that of 1 resolution beam.  Apart
from the binary, we only see three other relatively compact sources
within the field of view of $<10''$ (17,000 au) (Supplementary Figure
\ref{fig:lowmass}).  Two of these ($6''$ and $9''$ away from the
binary) may be protostars (panels b and c).  One other ($9''$ away
from the binary) does not have a point-like core, so we do not expect
it to be a protostellar source (panel d).  The masses of these
condensations are 0.022, 0.014, and 0.25 $M_\odot$, respectively,
assuming 30~K dust temperature.  In addition to these three sources,
there is also some emission to the south of the binary (panel e).
However, it appears to be extended and is part of the elongated
structure reaching to about $0.5''$ south of the binary seen in the
low-resolution data.  Therefore we also do not expect it to be a
protostar.  Thus our observations are sensitive to the presence of
low-mass protostellar sources, but we only see very limited number of
such sources in the region around the binary.  We consider that the
lack of other protostellar sources in the immediate environment of the
binary (i.e., the next closest source in projection is many ($\sim50$)
binary orbital separations away), is evidence against a turbulent
fragmentation scenario for formation of the binary.

\clearpage
\renewcommand{\figurename}{Supplementary Figure}
\setcounter{figure}{0}    

\begin{figure}
\begin{center}
\includegraphics[width=\textwidth]{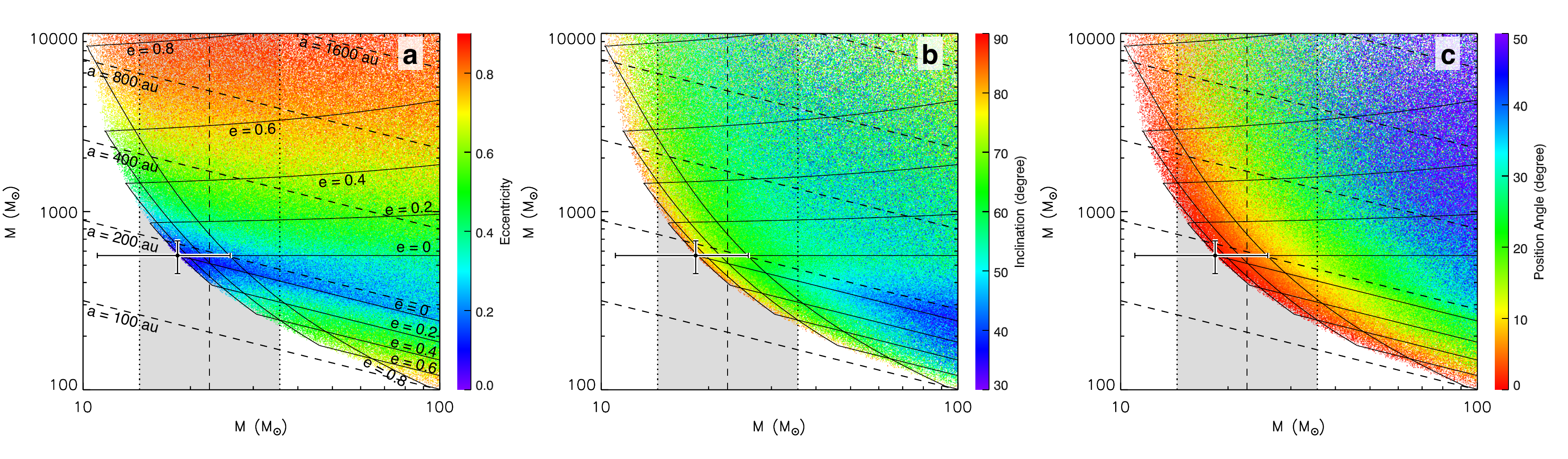}
\caption{
{\bf The distribution of the possible binary properties in the space
  of system mass and orbital period. a:} The color shows the
eccentricity of the orbits. {\bf b:} The color shows the inclination
of the orbital plane relative to the plane of sky. {\bf c:} The color
shows the position angle of the intersection line between the orbital
plane and the sky plane, with respect to the position angle of the
line connecting Sources A and B.  All possible elliptical orbits with
eccentricities from 0 to 0.9 are considered.  The data point and error
bars correspond to the system mass, orbital period, and their $1\sigma$
uncertainties, assuming an edge-on circular orbit with the apparent
separation of the two sources as their true separation.  The dashed
vertical line and the shaded regions indicate the system mass and its
uncertainties derived from the free-free emissions.  The dashed
diagonal lines show the locations of orbits with different semi-major
axes, as labelled in panel a.  The solid lines encircle the regions for
eccentricities of $e=0$, 0.2, 0.4, 0.6, and 0.8 (labelled in panel
a).}
\label{fig:binary1}
\end{center}
\end{figure}

\clearpage

\begin{figure}
\begin{center}
\includegraphics[width=\textwidth]{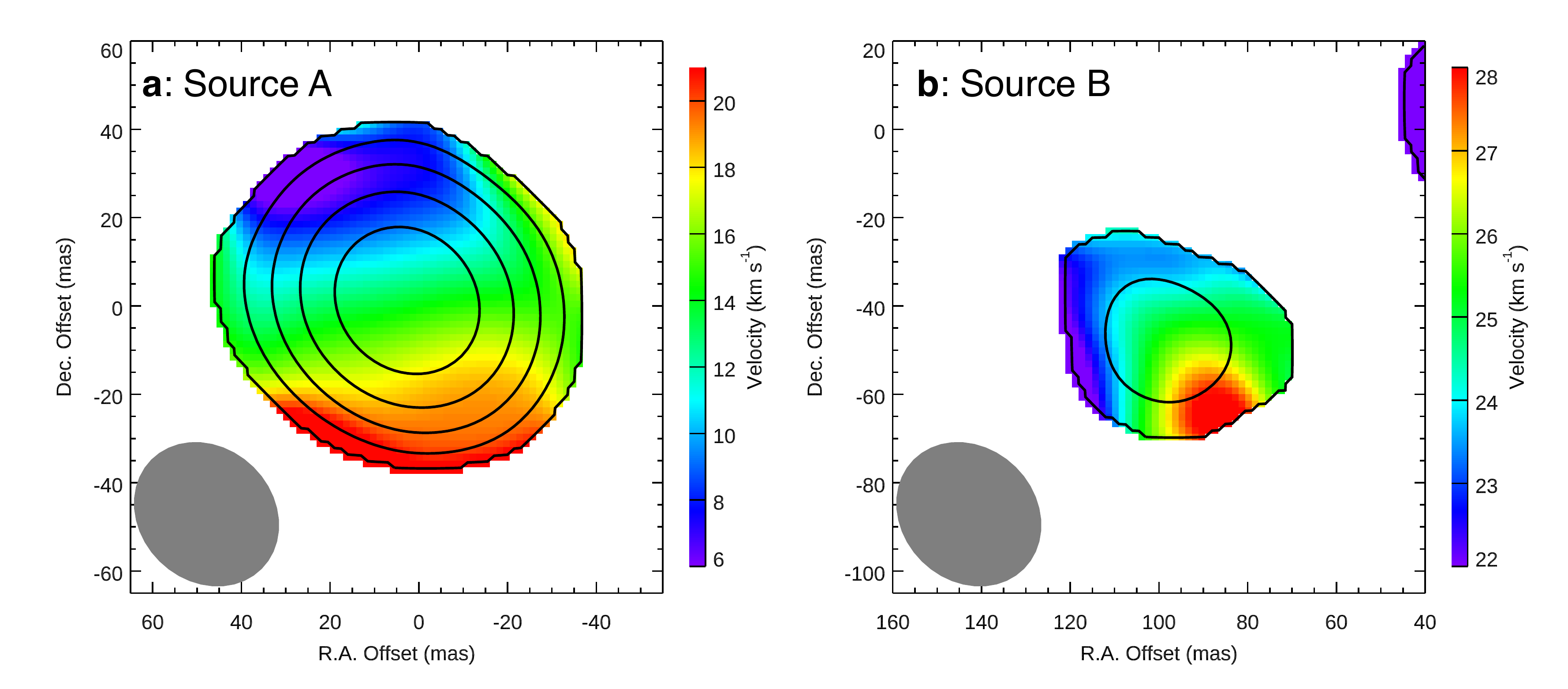}
\caption{
{\bf The moment maps of the H30$\alpha$ emission in Source A (panel a)
  and B (panel b).}  The moment 0 maps (integrated emission) are shown
in the black contours and the moment 1 maps (intensity-weighted mean
velocities) are shown in the color scale.  The moment 0 map in panel a
is integrated in the velocity range of $-30~\kms<\vlsr<55~\kms$, and
the contour levels are $5\sigma\times 2^n$ ($n=1,2,3,...$) with
$1\sigma=13~\mjybeam~\kms$.  The moment 0 map in panel b is integrated
in the velocity range of $-10~\kms<\vlsr<60~\kms$, and the contour
levels are $5\sigma\times 2^n$ ($n=1,2,3,...$) with
$1\sigma=14~\mjybeam~\kms$.  The synthesized beam is shown in the
bottom-left corner.}
\label{fig:momentmap}
\end{center}
\end{figure}

\clearpage

\begin{figure}
\begin{center}
\includegraphics[width=\textwidth]{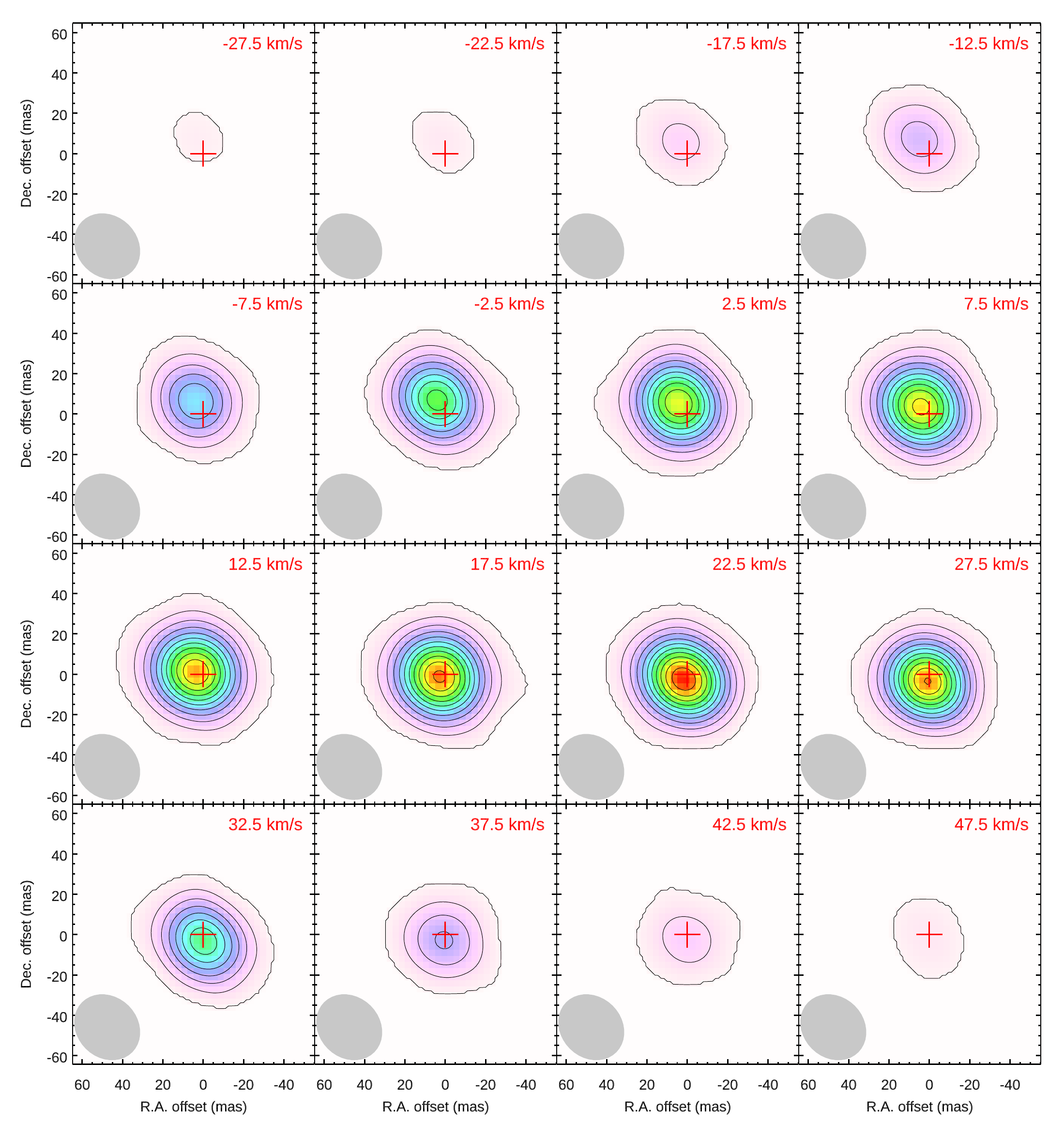}
\caption{{\bf The channel maps of the H30$\alpha$ emission in Source A.} The
channel width is $5~\kms$ and the central velocities of the channels
are labeled in the upper-right corners of the panels.  The contours
start at 5$\sigma$ and have an interval of 10$\sigma$ with
$1\sigma=17~\K$ (0.8 $\mjybeam$).  The synthesized beam is shown in
the bottom-left corner of each panel.  The red cross marks the
position of the continuum peak.}
\label{fig:channelmap}
\end{center}
\end{figure}

\clearpage

\begin{figure}
\begin{center}
\includegraphics[width=\textwidth]{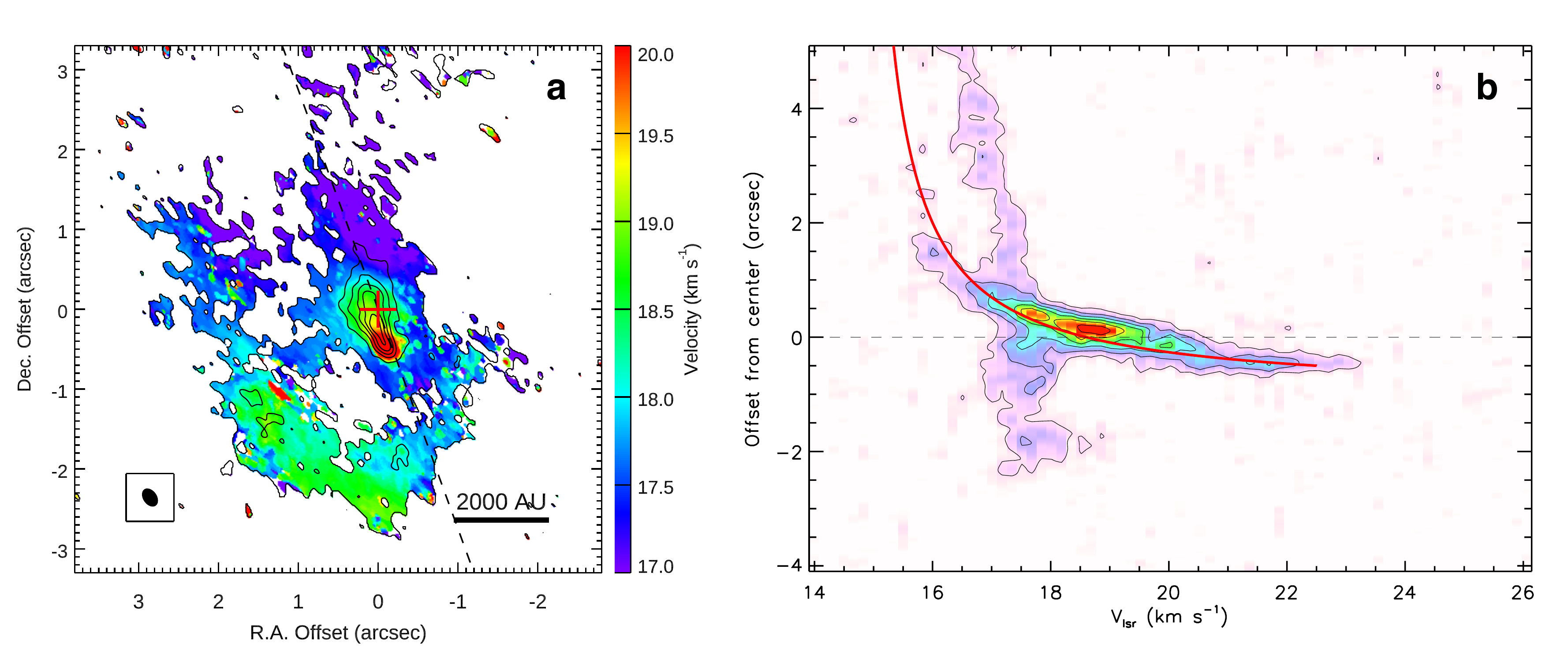}
\caption{
{\bf The kinematics of the large scale CH$_3$OH emission.}
{\bf a:} The moment maps of the large scale CH$_3$OH $4(2, 2)-3(1,
  2)$ emission.  The moment 0 map (emission integrated from
$\vlsr=15$ to $23~\kms$) is shown in the black contours and the moment
1 map (intensity-weighted mean velocities) is shown in the color
scale.  The contour levels start at 3$\sigma$ and have intervals of
6$\sigma$ ($1\sigma=7~\mjybeam~\kms$).  The synthesized beam, shown
inside the bottom-left square, is $0.25''\times0.17''$.  The red cross
marks the position of the continuum peak.  The dashed line indicates
the cut for the position-velocity diagram shown in panel b.  {\bf
  b:} The position-velocity diagram of the CH$_3$OH emission along
  the cut shown in panel a.  The cut width is $0.5''$.  The contour
levels start at 3$\sigma$ and have intervals of 3$\sigma$
($1\sigma=4~\mjybeam$).  The red curve is a model for explaining the
kinematics (see Supplementary Discussion).}
\label{fig:infall}
\end{center}
\end{figure}

\clearpage

\begin{figure}
\begin{center}
\includegraphics[width=\textwidth]{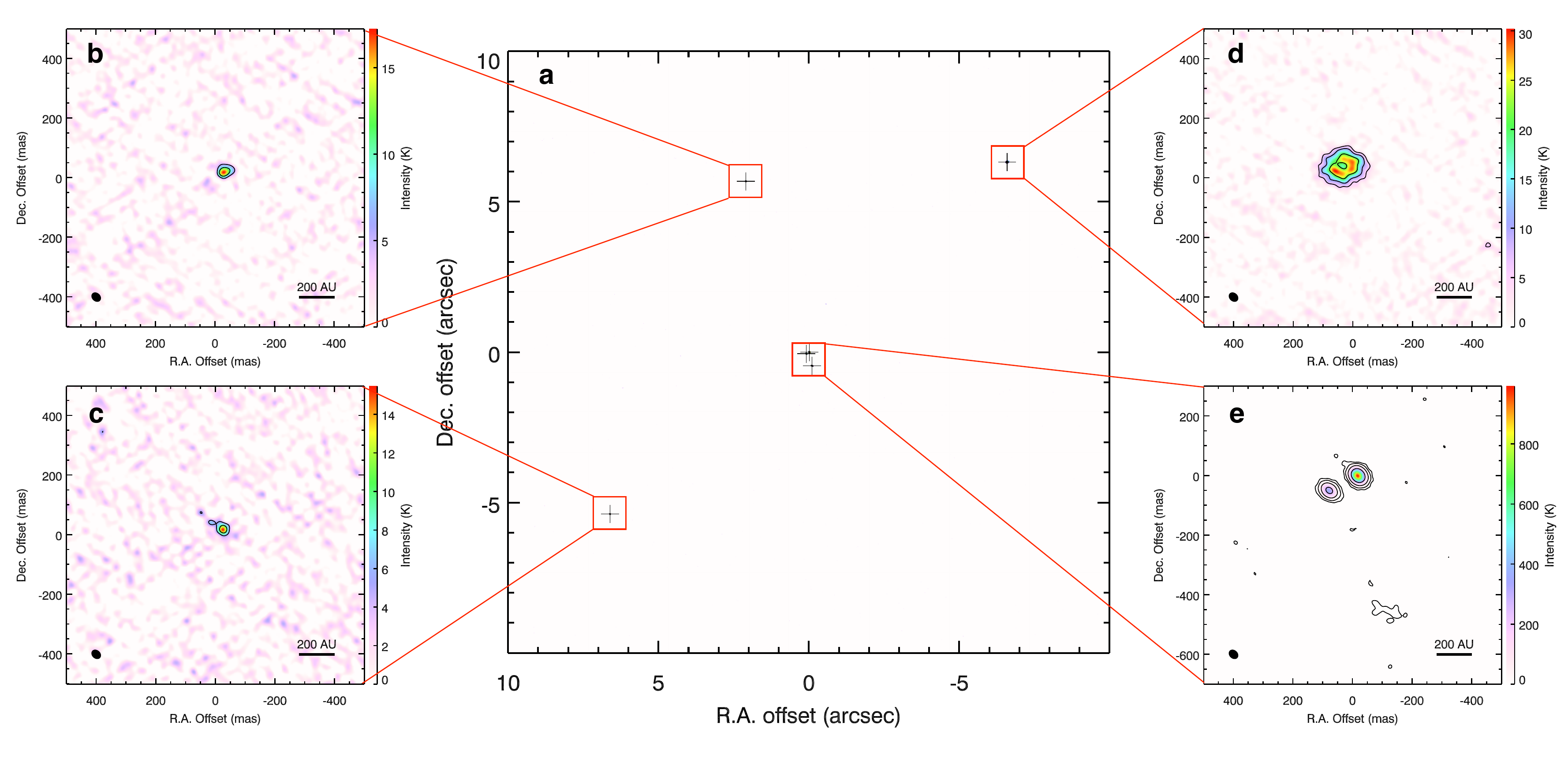}
\caption{{\bf Compact continuum sources identified in
the high-resolution data. a:} continuum map of the region $<10''$
from the central sources, with the crosses marking the identified 
continuum sources. {\bf b$-$e:} Zoom-in views of the identified continuum sources.}
\label{fig:lowmass}
\end{center}
\end{figure}

\clearpage

\begin{figure}
\begin{center}
\includegraphics[width=0.5\textwidth]{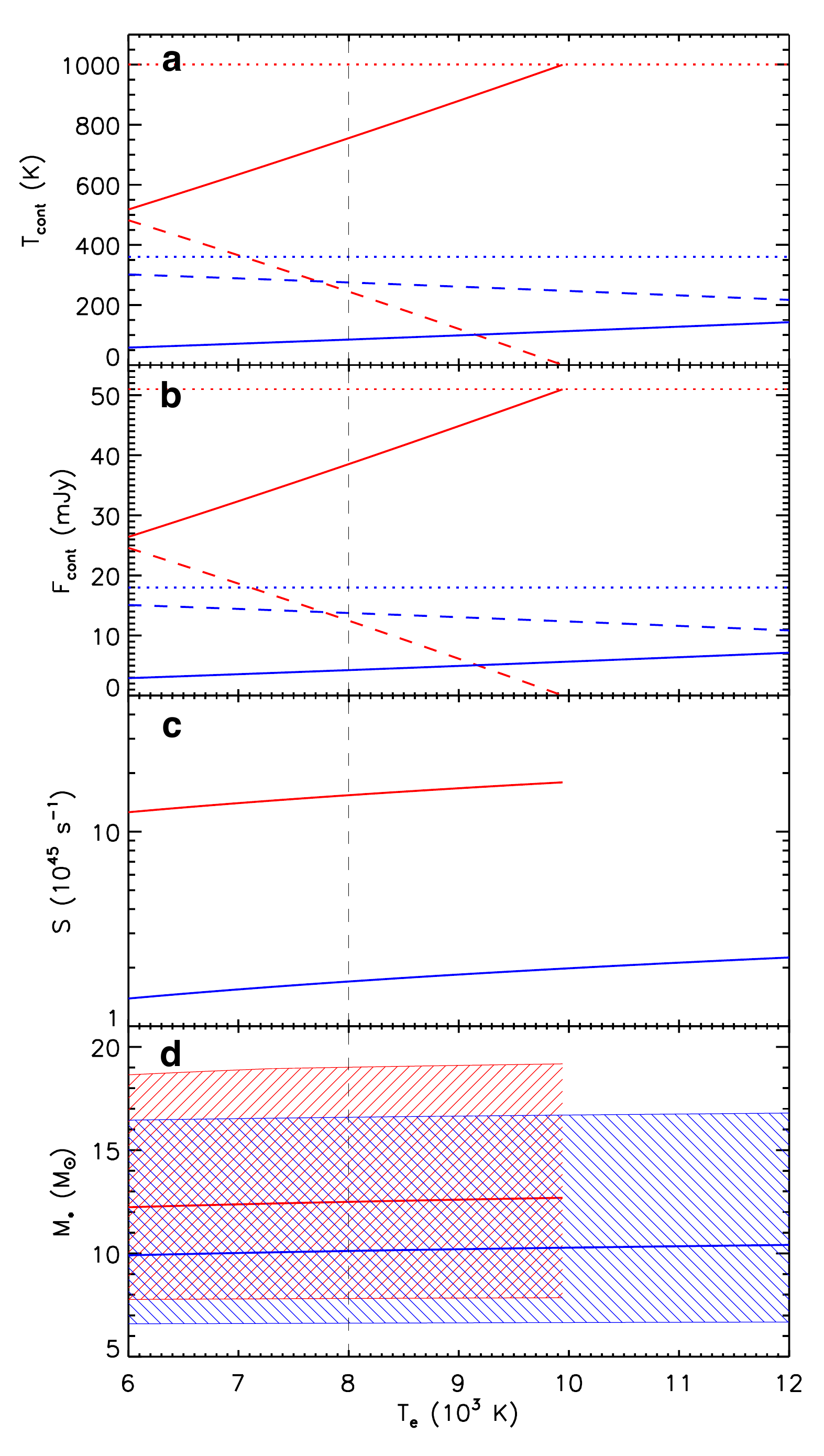}
\caption{{\bf Dependence of the estimated free-free peak intensities (panel a),
  fluxes (panel b), ionizing photon rates (panel c), and 
  protostellar masses (panel d) of the two sources on the
  assumed ionized gas temperature.} The red and blue colors are for
Source A and B, respectively. The solid, dashed, and dotted lines in
panels a and b are for the free-free, dust continuum, and total
continuum emissions.  The solid lines in panel d are estimated ZAMS
stellar masses, and the shaded regions indicate the ranges of
protostellar masses calculated from different accretion histories.
The vertical line indicates the temperature of 8000 K of ionized gas,
which is used as the fiducial case in the main text.}
\label{fig:freefree}
\end{center}
\end{figure}

\end{document}